\documentclass[sigconf,screen,authorversion]{acmart}
\AtBeginDocument{%
  \providecommand\BibTeX{{%
    \normalfont B\kern-0.5em{\scshape i\kern-0.25em b}\kern-0.8em\TeX}}}

\usepackage[nolist,nohyperlinks,smaller]{acronym}
\usepackage{siunitx}
\usepackage{calculator}
\usepackage{subcaption}
\usepackage{fmtcount}
\usepackage{csquotes}
\usepackage{cleveref}

\begin{acronym}[UMLX]
	\acro{HMD}{head-mounted display}
	\acro{GUI}{graphical user interface}
	\acro{HUD}{head-up display}
	\acro{TCT}{task-completion time}
	\acro{SVM}{support vector machine}
	\acro{EMM}{estimated marginal mean}
	\acro{AR}{Augmented Reality}
	\acro{VR}{Virtual Reality}
	\acro{RTLX}{Raw Nasa-TLX}
	\acro{WUI}{Walking User Interface}
	\acro{CSCW}{Computer-Supported Cooperative Work}
	\acro{HCI}{Human-Computer Interaction}
	\acro{ABI}{Around-Body Interaction}
	\acro{SLAM}{Simultaneous Location and Mapping}
	\acro{ROM}{range of motion}
	\acro{EMG}{electromyography}
	\acro{STS}{System Trust Scale}
	\acro{IPQ}{System Trust Scale}
	\acro{TLX}{NASA Task Load Index}
	\acro{ART}{Aligned Rank Transform}
	\acro{SSQ}{Simulator Sickness Questionnaire}
\end{acronym}

\copyrightyear{2023} 
\acmYear{2023} 
\setcopyright{acmlicensed}
\acmConference[CHI '23]{Proceedings of the 2023 CHI Conference on Human Factors in Computing Systems}{April 23--28, 2023}{Hamburg, Germany}
\acmBooktitle{Proceedings of the 2023 CHI Conference on Human Factors in Computing Systems (CHI '23), April 23--28, 2023, Hamburg, Germany}
\acmPrice{15.00}
\acmDOI{10.1145/3544548.3581557}
\acmISBN{978-1-4503-9421-5/23/04}

\newcommand{\systemname}{UndoPort}
\newcommand{\longtitle}{Exploring the Influence of Undo-Actions for Locomotion in Virtual Reality on the Efficiency, Spatial Understanding and User Experience}

\newcommand{\teleport}{point\&teleport}

\newcommand{\ivPositionUndo}{\textsc{position undo}}
\newcommand{\ivPositionUndoLvlYes}{\textsc{with position undo}}
\newcommand{\ivPositionUndoLvlNo}{\textsc{without position undo}}

\newcommand{\ivOrientationUndo}{\textsc{orientation undo}}
\newcommand{\ivOrientationUndoLvlYes}{\textsc{with orientation undo}}
\newcommand{\ivOrientationUndoLvlNo}{\textsc{without orientation undo}}

\newcommand{\ivMovementVisualization}{\textsc{movement visualization}}
\newcommand{\ivMovementVisualizationLvlCont}{\textsc{continuous}}
\newcommand{\ivMovementVisualizationLvlDisc}{\textsc{discrete}}

\newcommand{\dvCollectionTime}{\textsc{coin collection time}}
\newcommand{\dvNrOfTeleports}{\textsc{number of teleports}}
\newcommand{\dvTotalMovementActions}{\textsc{number of movement actions}}
\newcommand{\dvDistance}{\textsc{traveled distance}}

\newcommand{\dvTimeBeforeFirstCorridor}{\textsc{time before first corridor}}

\newcommand{\dvRevisitError}{\textsc{revisit error}}

\newcommand{\dvTLX}{\textsc{\acs{TLX}}}
\newcommand{\dvSSQ}{\textsc{\acs{SSQ}}}
\newcommand{\dvLikert}{\textsc{custom questionnaire}}
\newcommand{\dvPresence}{\textsc{presence}}

\newcommand{\experimentPhaseOne}{training phase}
\newcommand{\experimentPhaseTwo}{experimental phase}

\newcommand{\repetions}{2}
\newcommand{\coinsInExperimentPhase}{4}

\newcommand{\ano}[4]{$F_{#1, #2}=#3$, $p#4$}

\newcommand{\subEtaG}[2]{%
	\ifthenelse{\equal{#1}{\string >.05}}
	{}
	{, $\eta_{G}^{2}=#2$}%
}

\newcommand{\subEta}[2]{%
	\ifthenelse{\equal{#1}{\string >.05}}
	{}
	{, $\eta^{2}=#2$}%
}

\newcommand{\chisq}[3]{$\chi^2(#1) = #2$, $p #3$}

\newcommand{\valSi}[3]{$M = \SI[round-mode=places,round-precision=1]{#1}{#3}$, $SD = \SI[round-mode=places,round-precision=1]{#2}{#3}$}
\newcommand{\val}[2]{$M~=~\num[round-mode=places,round-precision=1]{#1}$, $SD~=~\num[round-mode=places,round-precision=1]{#2}$}

\def\gge{$\epsilon$}

\def\ges{$\eta_{G}^{2}$}

\newcommand{\efETAsquared}[1]{%
	\ifdim#1pt>0.139pt 
	large (\ges{} = #1)
	\else 
	\ifdim#1pt>0.059pt 
	medium (\ges{} = #1)
	\else 
	small (\ges{} = #1)
	\fi
	\fi
}

\usepackage{etoolbox}
\newcommand{\pquote}[2]{\enquote{\emph{#1}} (#2)}

\AtBeginMaketitle{%
\let\oldAtBeginDocument\AtBeginDocument%
\renewcommand\AtBeginDocument[1]{#1}
\setcopyright{acmlicensed}
\copyrightyear{2023}
\acmYear{2023}
\acmDOI{10.1145/3544548.3581557}
\acmISBN{978-1-4503-9421-5/23/04}
\let\AtBeginDocument\oldAtBeginDocument%
}

\begin{document}

\title[\systemname{}]{\systemname{}: \longtitle{}}

\author{Florian Müller}
\orcid{0000-0002-9621-6214}
\affiliation{%
	\institution{LMU Munich}
	\city{Munich}
	\country{Germany}
}
\email{florian.mueller@ifi.lmu.de}

\author{Arantxa Ye}
\orcid{0000-0002-0708-2817}
\affiliation{%
	\institution{LMU Munich}
	\city{Munich}
	\country{Germany}
}
\email{A.Ye@campus.lmu.de}

\author{Dominik Schön}
\orcid{0000-0003-2704-2852}
\affiliation{%
	\institution{TU Darmstadt}
	\city{Darmstadt}
	\country{Germany}
}
\email{schoen@tk.tu-darmstadt.de}

\author{Julian Rasch}
\orcid{0000-0002-9981-6952}
\affiliation{%
	\institution{LMU Munich}
	\city{Munich}
	\country{Germany}
}
\email{julian.rasch@ifi.lmu.de}

\renewcommand{\shortauthors}{Müller et al.}

\begin{abstract}

When we get lost in Virtual Reality (VR) or want to return to a previous location, we use the same methods of locomotion for the way back as for the way forward. This is time-consuming and requires additional physical orientation changes, increasing the risk of getting tangled in the headsets' cables. In this paper, we propose the use of undo actions to revert locomotion steps in VR. We explore eight different variations of undo actions as extensions of point\&teleport, based on the possibility to undo position and orientation changes together with two different visualizations of the undo step (discrete and continuous). We contribute the results of a controlled experiment with 24 participants investigating the efficiency and orientation of the undo techniques in a radial maze task. We found that the combination of position and orientation undo together with a discrete visualization resulted in the highest efficiency without increasing orientation errors.

\end{abstract}



\begin{teaserfigure}
	\includegraphics[width=\textwidth]{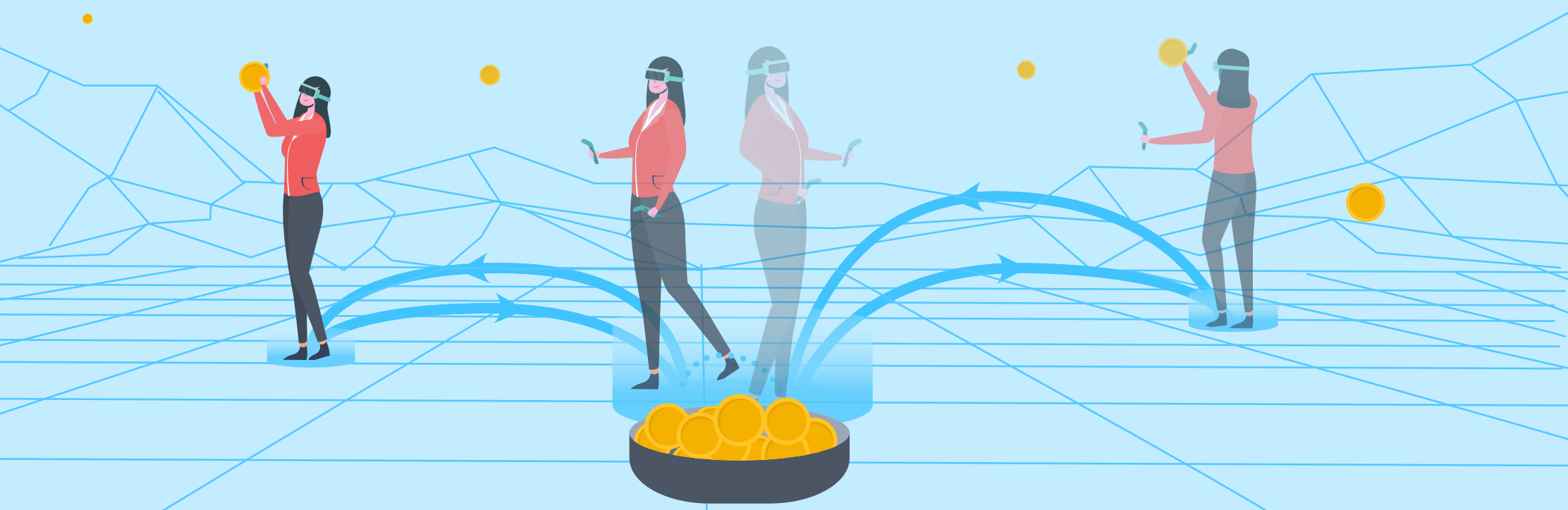}
	\caption{We present UndoPort, an extension of the \teleport{} locomotion technique with undo actions. UndoPort allows users to revert changes to their position and orientation and, thus, allows users to jump back to previously visited waypoints. In this work, we evaluate undo actions in terms of their impact on efficiency, local understanding, and user experience.}
	\Description[An illustration of undo-actions to revert locomotion steps in VR.]{An illustration of point\&click teleport locomotion in VR. After jumping to a target, the user can undo the locomotion step to go back to a prior position.}
	\label{fig:undoport:teaser}
\end{teaserfigure}

\maketitle

\section{Introduction}
\label{sec:introduction}

When exploring unfamiliar territory or collecting items in our known surroundings, we are often faced with the need to retrace paths to get to previous waypoints, such as junctions or a central starting point. Similar to the real world, this backtracking of known routes with the goal of reaching a previously visited waypoint is particularly common in \ac{VR}, where exploration tasks such as finding~\cite{Funk2019} or collecting~\cite{vanGelder2017} items or information are essential mechanics in gaming~\cite{Bovim2021} and learning environments~\cite{Pan2006, Shi2021b, Marky2019}. In familiar environments, this repetitive traversal of known locations reduces exploration efficiency. In unfamiliar environments, difficulty relocating places recently visited~\cite{Darken1995} can additionally lead to disorientation, lower performance, and spatial knowledge acquisition~\cite{Darken1996}.

While this \emph{going back} is a necessity to reach previously visited locations in reality, locomotion in \ac{VR} is not subject to the physical laws of reality. From point\&click teleport~\cite{Bozgeyikli2016} or walk-in-place techniques~\cite{Templeman1999} to foot movements~\cite{Iwata2006, Willich2020} or weight shifting in chairs~\cite{VonWillich2019}, research and industry have proposed a plethora of artificial locomotion techniques to address the mismatch between the limited size of the physical tracking space and the potentially boundless vastness of virtual worlds. While practical and valuable for exploring \ac{VR} environments, we still employ the same method of locomotion to return to a previous waypoint, just as we would in reality.

In this paper, we go beyond state-of-the-art and add to the body of research in \ac{VR} locomotion techniques by exploring undo-actions for locomotion in \ac{VR} to quickly return to previous waypoints. For this, we propose to record the user's locomotion history and allow them to jump back to any previous waypoint by pressing a button (see \cref{fig:undoport:teaser}). We explore the proposed undo concept as an extension of point-and-click teleport, which we chose as a baseline due to its status as the de-facto standard for locomotion in \ac{VR} in industry, as well as the inherent existence of waypoints.

The contribution of this paper is two-fold. First, we contribute the results of a controlled experiment assessing the influence of undo actions on efficiency, spatial understanding, and user experience in a \ac{VR} maze task. Here, we investigated eight different implementations of such an undo concept based on the possibility of undoing 1) position and 2) orientation changes together with 3) two different visualizations of the undo step (discrete and continuous). Second, based on the results of the controlled experiment, we contribute a set of guidelines and lessons learned for the future usage of undo actions to support locomotion in \ac{VR}.

\section{Related Work}
\label{sec:relatedwork}

A large body of prior work on 1) locomotion techniques for virtual reality heavily influenced our work. In the following section, we discuss these works with an in-depth focus on 2) point\&click locomotion techniques. 

\subsection{Locomotion in Virtual Reality}
\label{sec:relatedwork:locomotion}

While virtual worlds are only constrained in their spatial dimensions by the designer's imagination, the tracking space in the physical world is not. This mismatch limits the suitability of natural human motion as a means of locomotion in \ac{VR} to room-scale-based virtual environments~\cite{Langbehn2018}. As a solution to this mismatch between the limited size of the tracking area and the potentially unlimited virtual worlds, research has proposed a wide variety of artificial locomotion methods that decouple movement in the physical (tracked) world from movement in the virtual world. Locomotion Vault\footnote{\url{https://locomotionvault.github.io/}}~\cite{DiLuca2021} provides a comprehensive overview of locomotion techniques. Many different classifications and categorizations exist for such artificial locomotion techniques for \ac{VR} in the literature. However, a central criterion of distinction is typically the classification into 1) continuous or 2) discrete locomotion techniques~\cite{Boletsis2017, AlZayer2020}.

Continuous locomotion techniques visually resemble the way we are experiencing locomotion in the physical world by applying changes in translation in the virtual scene over time~\cite{Bowman1997}, completely decoupling virtual locomotion from the translation of the user's body in the physical world. Such techniques leverage controllers ~\cite{Englmeier2020} or other accessories like chairs~\cite{VonWillich2019, Rietzler2018, Gugenheimer2016, Nguyen-Vo2021} or shoes~\cite{Liu2021}. Further, research also proposed to leverage head~\cite{Tregillus2017} or hand gestures~\cite{Ferracani2016,Schafer2021, Cardoso2016}. As another possible solution, techniques like treadmills~\cite{Cakmak2014}, in-place~\cite{Lee2019, Kreimeier2019}, scaled~\cite{Abtahi2019, Wilson2018} or redirected walking~\cite{Nilsson2018} alter the user's visual perception to allow for unconstrained continuous movement in the virtual world while walking on-spot or in small circles in the physical world. In recent years, research has expanded such continuous locomotion techniques from 2D to 3D~\cite{Rheiner2014, Sasaki2019, Chen2013, Zhang2019} environments. While practical and valuable, continuous locomotion techniques are known to be prone to cybersickness~\cite{Mayor2021} or require larger tracking areas~\cite{Englmeier2020}.

Research proposed discretizing the target selection and locomotion process to overcome these limitations of continuous locomotion techniques. As the most prominent example, teleportation techniques such as \teleport{}~\cite{Bozgeyikli2016, Funk2019}, portals~\cite{Freitag2014}, or fixed nodes~\cite{JacobHabgood2018} allow users to skip the movement but directly jump to (intermediate) target locations. Research has shown that such discrete locomotion techniques allow for fast~\cite{Mayor2021} and accurate~\cite{Funk2019} travel while lowering the problem of cybersickness~\cite{JacobHabgood2018}. However, research showed that the visual jumps could break the users' sense of presence~\cite{Mayor2021} and decrease spatial awareness~\cite{Bowman1997}, diminishing their usefulness in certain situations. 

Considering the discussed advantages and disadvantages of continuous and discrete locomotion techniques, we opted to build our proposed technique on top of \teleport{}, the most prominent discrete locomotion technique. In the following section, we present a more in-depth discussion of \teleport{}. For a more detailed classification of general \ac{VR} locomotion techniques, we refer to the excellent works of \citet{Boletsis2017} and \citet{AlZayer2020}. 

\subsection{Point\&Click Teleport}

\begin{figure*}[ht!]
	\subfloat[Radial Maze Task (Training Phase)\label{fig:undoport:maze:maze}]
	{\includegraphics[width=.33\linewidth]{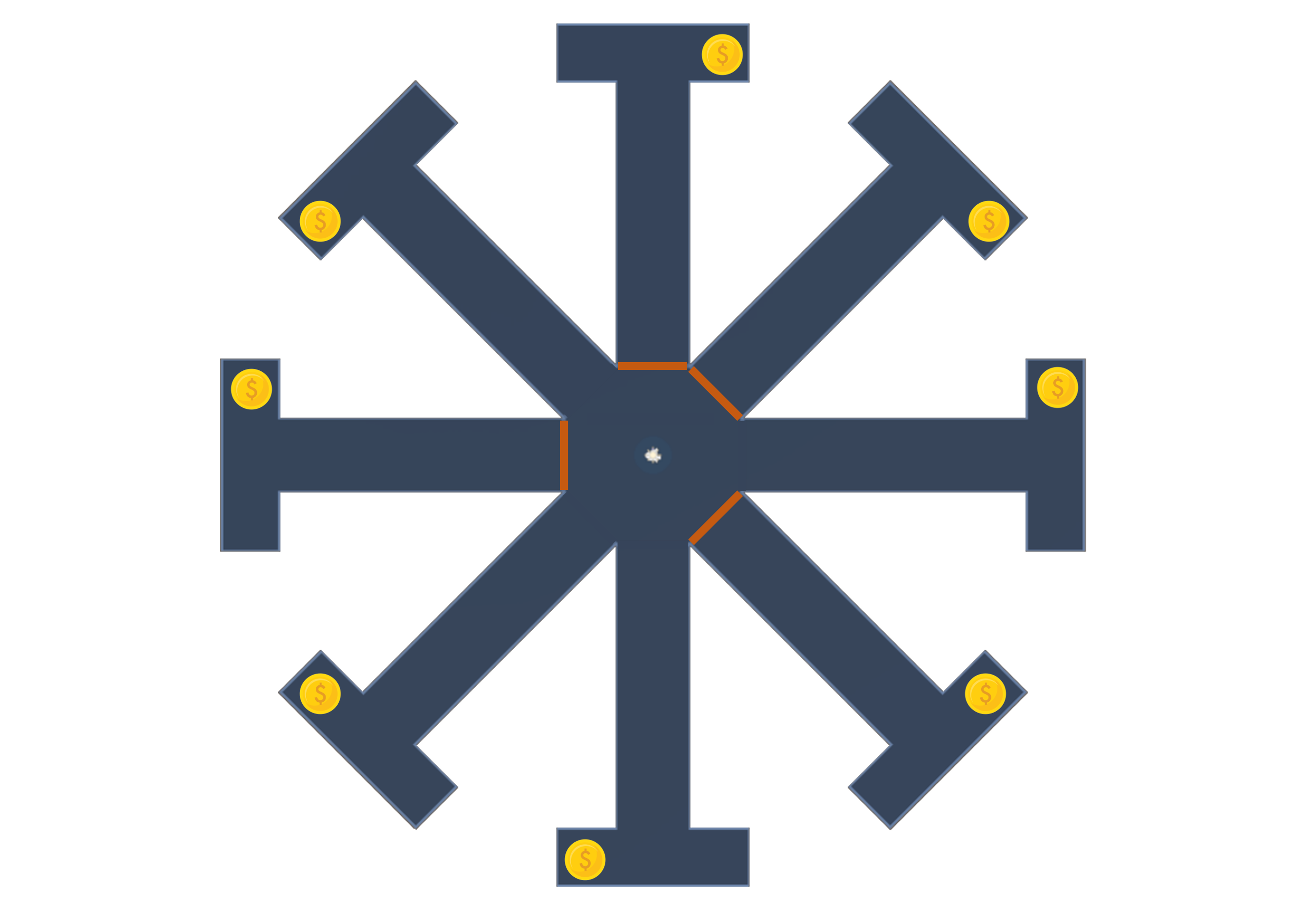}}\hfill
	\subfloat[Radial Maze Task (Experiment Phase)\label{fig:undoport:maze:training}]
	{\includegraphics[width=.33\linewidth]{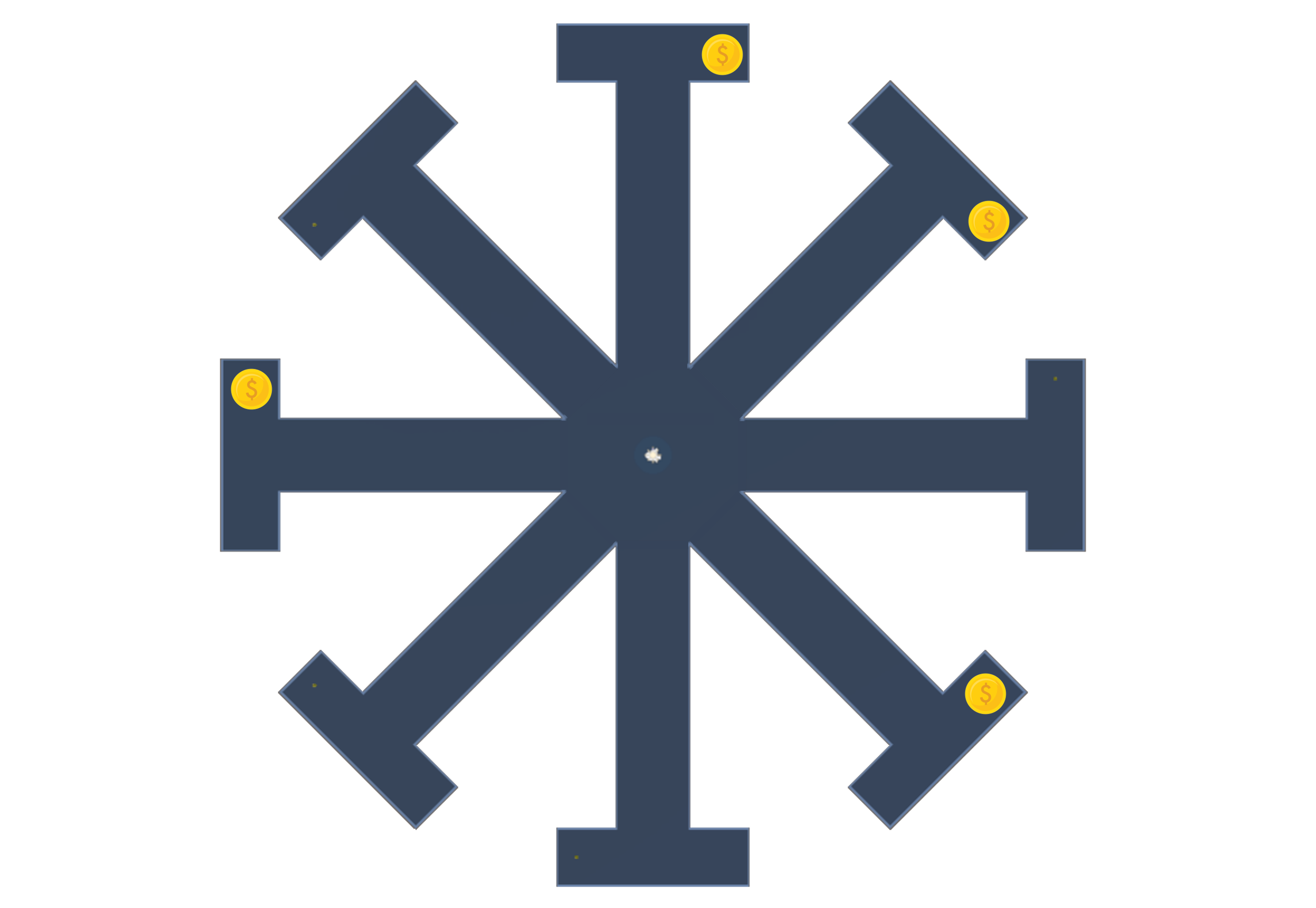}}\hfill
	\subfloat[Study Environment\label{fig:undoport:maze:experiment}]
	{\includegraphics[width=.33\linewidth]{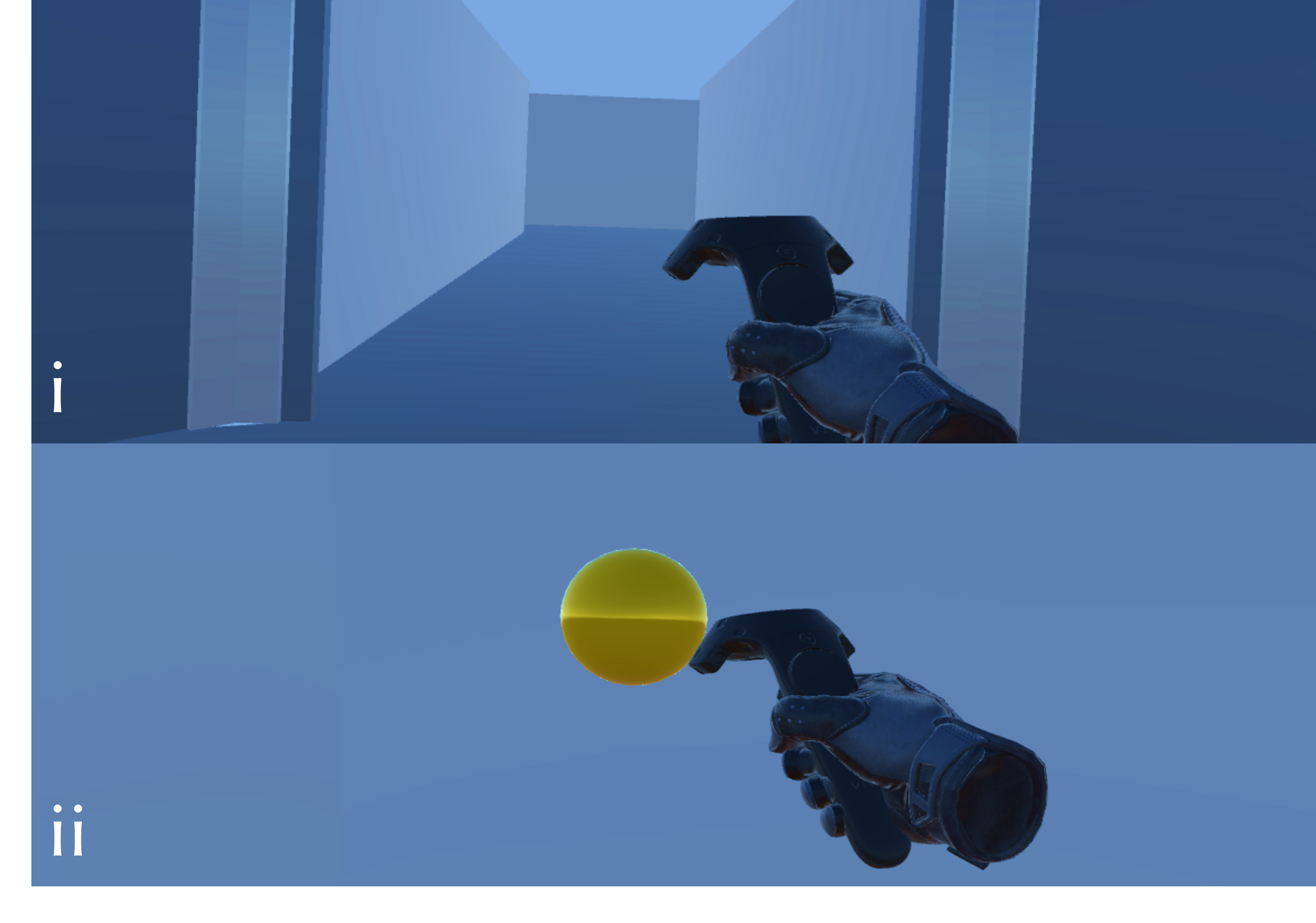}}\hfill
	\caption{The radial maze task used in the controlled experiment. During the (a) \experimentPhaseOne{}, the participants' task was to collect the coins from the initially open four coridors. After participants had collected the last coin, the remaining four coridors opened for the (b) \experimentPhaseTwo{}. Again, the participants' task was to collect the remaining four coins by (c) teleporting through the maze and pressing a button close to the coins.}
	\Description[An illustration of the radial maze task used in the experiment.]{A three-part figure depicting a radial maze with eight corridors from a birds-view perspective (a,b). In the training phase (a), four of the corridors are closed. In the experiment phase (b), four of the coins have been collected, and the remaining four doors are open. Subfigure (c) depicts screenshots from the participants' perspective (i) into one of the corridors and (ii) while grabbing a coin.}
\end{figure*}

As the most prominent example of a discrete locomotion technique, \teleport{} has gained substantial interest from the research community and has become the de-facto standard in commercial \ac{VR} games. While \citet{Bowman1997} already explored pointing-based locomotion techniques in 1997 and others further explored the topic~\cite{Bolte2011, Freitag2014}, \citet{Bozgeyikli2016} first introduced the name and compared \teleport{} to walk-in-place and joystick-based locomotion.

In recent years, research proposed a variety of extensions and modifications to \teleport{}. \citet{Funk2019} and \citet{Bozgeyikli2016} explored adjusting the users' orientation during the aiming phase. Further, research explored other body parts, such as eye~\cite{Kth2017} or head gaze~\cite{Christou2017} and foot movements~\cite{Willich2020, Carrozzino2014} to select the target. Finally, \citet{Matviienko2022} extended \teleport{} to 3D locomotion by enabling users to cut off the ray and \citet{Weissker2019} and \citet{raschGoingGoingGone2023} extended \teleport{} for joint multi-user locomotion.

Further, research proposed various solutions to overcome users' spatial understanding and orientation problems. \citet{Cmentowski2019} and \citet{Griffin2019} explored a third-person view for \teleport{}. Further, \citet{Xu2017} compared \teleport{} to joystick and walk-in-place locomotion and did not find significant differences regarding the spatial understanding of users. As a promising solution, \citet{Bhandari2018} proposed quickly and continuously moving the user to the target location instead of fading the users' view in and out.

While practical and valuable, today's \teleport{} techniques require us to physically turn around and use the same locomotion technique to return to previously visited waypoints. This process is time-consuming and can lead to tangling in cables~\cite{Funk2019}. As a possible solution, we explore undo-actions to allow users to return to previous waypoints without the need to rotate physically. To the best of our knowledge, there exists no prior literature explicitly focusing on returning to previously visited waypoints to backtrack the last steps of the locomotion. Following the promising results of \citet{Bhandari2018}, we further included the use of undo-actions for both types of motion visualization.

\section{Methodology}
\label{sec:methodology}

We conducted a controlled experiment to investigate the accuracy, efficiency, and user experience of undo-actions for locomotion actions as an addition to point\&click teleport as today's de-facto standard \ac{VR} locomotion technique. More specifically, we investigated the following research questions:

\begin{description}
	\item[RQ1] How does the ability to reset the position change of a locomotion action influence the accuracy, efficiency, and user experience of locomotion in \ac{VR}? 
	\item[RQ2] How does the ability to reset the orientation change of a locomotion action influence the accuracy, efficiency, and user experience of locomotion in \ac{VR}? 
	\item[RQ3] How does the ability to reset both the change in position and orientation of a locomotion action influence the accuracy, efficiency, and user experience of locomotion in \ac{VR}? 
\end{description}

\subsection{Design and Task}

We designed a controlled experiment in which participants used varying combinations of position- and orientation-undo for locomotion in a \ac{VR} maze task. To explore users' performance in terms of efficiency while also accounting for potential negative influences of the proposed techniques on participants' spatial understanding and memory, we used an adapted 8-arm radial maze task.

\subsubsection{Radial Arm Task}

\begin{figure*}[h!]
	\includegraphics[width=\linewidth]{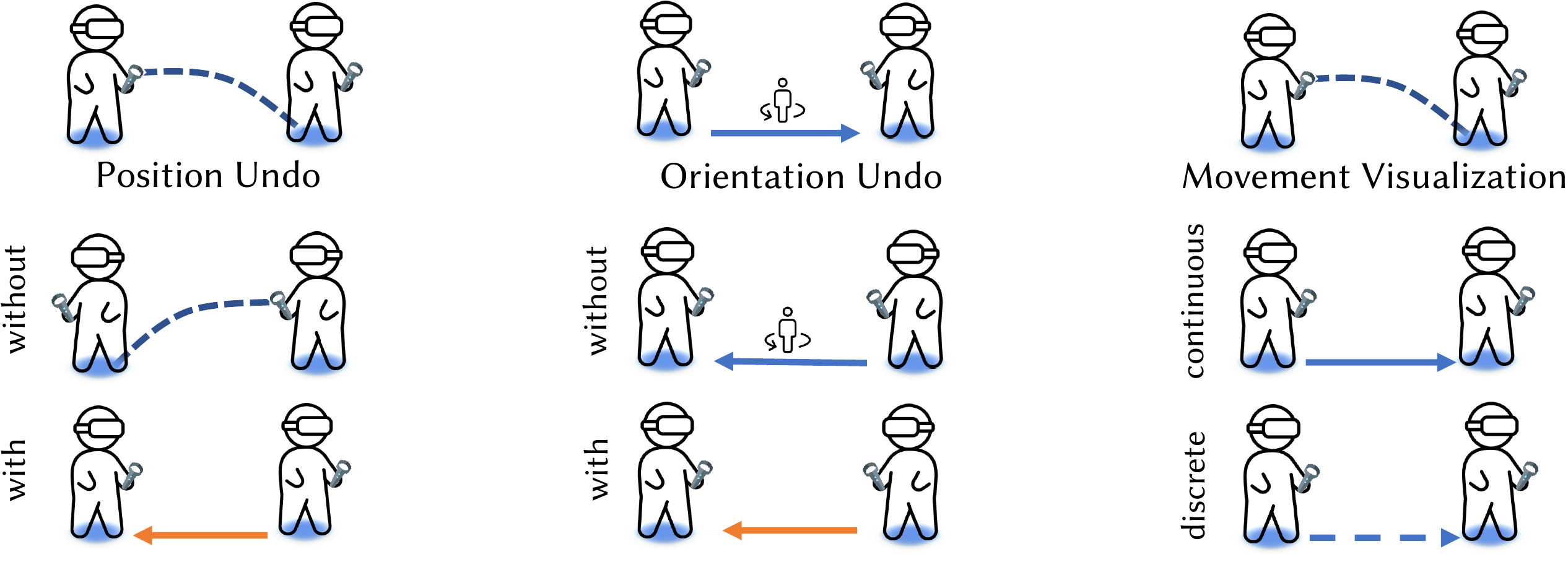}
	\caption{The independent variables studied in the experiment with their respective levels. From left to right: \ivPositionUndo{} (\ivPositionUndoLvlNo{} and \ivPositionUndoLvlYes{}), \ivOrientationUndo{} (\ivOrientationUndoLvlNo{} and \ivOrientationUndoLvlYes{}) and \ivMovementVisualization{} (\ivMovementVisualizationLvlCont{} and \ivMovementVisualizationLvlDisc{}).}
	\Description[The independent variables varied in the experiment.]{A three-part figure depicting the three independent variables varied in the experiment using abstract figures.}
	\label{fig:undoport:ivs}
\end{figure*}

The radial arm maze task was first used to assess the spatial abilities of rodents by \citet{Olton1976} in 1976. Since then, the task has been adapted for use with humans in real~\cite{Olton1987, Spieker2012, Oconnor1993} and virtual~\cite{Kim2018, BenZeev2020} settings. The basic version of the task consists of a central room, from which a certain number (usually 8) uniform corridors spread. At the end of the corridors, there are hidden rewards that the test subject is supposed to reach. There are a variety of variations of the radial arm task in the literature, which vary in the exact task, the number of arms, and the amount of external information through visual cues in the world. Further, the literature distinguishes radial maze tasks between free-choice and forced-choice variants, depending on whether all arms are open at the beginning (free-choice) or whether a specific subset of the arms must be visited first (forced-choice)~\cite{Palombi2022}. 

To exclude the influence of external visual cues and prevent the strategic circular progression, we adapted an uncued forced-choice radial maze task as follows: From the central room, 8 uniform corridors depart, each, in turn, branching at the end in a T-junction. The central room is connected to each corridor through a door  (see \cref{fig:undoport:maze:maze}). A coin is hidden in one of the two T corridors for each corridor. The coin is not visible from the central room. At the beginning of each trial, the participants are placed in the center of the room. The trial now consists of two phases:

\begin{description}
	\item[\experimentPhaseOne{}] In the first phase, 4 randomly selected corridors are accessible while the doors leading to the other corridors are closed (see \cref{fig:undoport:maze:training}). The participant's task is to collect the coins from the corridors as quickly as possible by pulling the trigger while the controller is in close proximity to the coin.
	\item[\experimentPhaseTwo{}]  After the last of the 4 directly accessible coins is collected, the four other doors open. In the second phase, the participants' task is to collect the remaining 4 coins (see \cref{fig:undoport:maze:experiment}).
\end{description}

We opted for this variation of the radial maze task because, by taking out the freedom to explore the arms at will, this variant shifts the focus away from search strategies toward spatial understanding and memory. Further, this particular version of a maze allows the generation of comparable yet different tasks over multiple repetitions.

\subsubsection{Independent Variables}

To assess a broad picture of the possible factors influencing efficiency, spatial understanding, and user experience of the interaction, we varied 3 independent variables:

\begin{description}
	\item[\ivPositionUndo{}] Following our general idea, we varied the ability to undo a locomotion step between \ivPositionUndoLvlYes{} and \ivPositionUndoLvlNo{} as our first independent variable. In the \ivPositionUndoLvlYes{} conditions, pressing the action button one time would teleport the participant to the last waypoint. Repeated usage of the action traces the participant's movement path further back, one waypoint at a time.
	\item[\ivOrientationUndo{}] Considering the literature review, we expected that the handling of the user's orientation during the reset would impact the performance parameters. Therefore, we varied the ability to undo orientation changes between \ivOrientationUndoLvlYes{} and \ivOrientationUndoLvlNo{} as the second independent variable. In the \ivOrientationUndoLvlYes{} conditions, pressing the action button resets the participant's orientation to the orientation captured at the beginning of the last teleport. More precisely, the keypress resets the orientation based on the participant's line of sight (that is, the orientation of the \ac{HMD}). As with the \ivPositionUndo{}, repeated usage of the action traces back to the previous waypoints of the participant.
	\item[\ivMovementVisualization{}] We hypothesized that undo actions could result in reduced spatial orientation. As a possible solution, we varied the visualization of movement between \ivMovementVisualizationLvlDisc{} and \ivMovementVisualizationLvlCont{} as a third independent variable. In \ivMovementVisualizationLvlDisc{} visualization, the user's view is faded to black and then faded back in at the new position, resulting in no visual cues about the traveled path. This is the default visualization for teleport techniques in use today. On the other hand, the \ivMovementVisualizationLvlCont{} visualization quickly changes the user's viewpoint over time and thus provides a visual flow during the movement, as proposed by \citet{Bhandari2018}. The authors demonstrated that this visualization can help to reduce spatial disorientation in teleport-based locomotion. To keep the conditions comparable, we used the respective visualization for all types of movement, i.e., for regular (forward) teleportations and undo actions for both position and orientation changes.
\end{description}

We varied our independent variables in a repeated-measures design, resulting in a total of $2 \times 2 \times 2=8$~conditions. In each condition, the participants performed the task as described above two times. As we only evaluate the four coins in the \experimentPhaseTwo{}, this yielded a total of $8 \times \coinsInExperimentPhase \times \repetions = 64$ trials per participant. To avoid learning effects, we counterbalanced the order of conditions in a balanced Latin square design with 8 levels. In addition, we chose a random distribution of initially closed corridors for \experimentPhaseOne{} and randomized the coin's position in the left or right T arm.

\subsubsection{Dependent Variables}

To answer our research questions, we logged the following dependent variables for each trial in the \experimentPhaseTwo{}.

\begin{description}
	\item[\dvCollectionTime{}] as the time (in \si{\second}) required to collect the coin. We started the timer with the collection of the previous coin. 
	\item[\dvNrOfTeleports{}] as the number of teleports (forward) used to reach the coin. 
	\item[\dvTotalMovementActions{}] as the total number of movement actions (teleport and undo) used to reach a coin.
	\item[\dvDistance{}] as the traveled distance (in \si{\meter}) to reach the coin.
	\item[\dvTimeBeforeFirstCorridor{}] as the time (in \si{\second}) between starting \experimentPhaseOne{} and the participant entering the first corridor.
	\item[\dvRevisitError{}] as the number of visits to corridors that were already visited before. We counted the visit to a corridor as soon as the participant's position crossed the door threshold.
\end{description}

We reset all measurements when collecting a coin. Thus, all measurements refer to the path from one coin to the next, i.e., from the end of one corridor to the end of another. This includes the first coin of \experimentPhaseTwo{} since it was preceded by the last coin of \experimentPhaseOne{}. We only analyzed the four coins from \experimentPhaseTwo{}. In addition, after each condition, we asked participants to complete a questionnaire that included the following.

\begin{description}
	\item[\dvTLX{}] as the \acl{TLX} questionnaire as proposed by \citet{Hart1988} to assess the perceived workload of participants.
	\item[\dvSSQ{}] as the \acl{SSQ} questionnaire as proposed by \citet{Kennedy1993} to assess sickness induced by our interaction techniques.
	\item[\dvPresence{}] as the participants' self-assessment for their feeling of presence. For this, participants answered the question \enquote{In the computer-generated world I had a sense of \enquote{being there}} on a 7-point Likert scale (\enquote{not at all} $\ldots$ \enquote{very much}) as proposed by \citet{Slater1994}.
	\item[\dvLikert{}] Additionally, we asked the participants to answer questions on a 5-point Likert scale, assessing their user experience.
\end{description}

\subsection{Study Setup and Apparatus}

We implemented the radial maze using Unity 2021.3.4f1. The central room was round with a diameter of \SI{10}{\meter}. Each of the eight corridors was \SI{15}{\meter} long and \SI{3.8}{\meter} wide. Further, each corridor branched at the end with a T-junction (angle \SI{\pm 90}{\degree}) into two corridors, each \SI{5}{\meter} long and \SI{3.8}{\meter} wide. The room was \SI{3.5}{\meter} heigh and open to the top. The corridors were arranged in a circular pattern around the central room with relative angles of \SI{\pm 45}{\degree} (see \cref{fig:undoport:maze:maze,fig:undoport:maze:training,fig:undoport:maze:experiment}). The room layout provided no visual cues to the participant's current orientation. We visualized the rewards as spherical coins with a diameter of \SI{0.2}{\meter} floating at the participants' shoulder height of around \SI{1.4}{\meter} and \SI{1}{\meter} away from the end of the T arms. Participants collected coins by pressing the trigger button in close physical proximity to a coin (\SI{0.2}{\meter}). In addition to the coin disappearing, we added an auditory signal communicating the successful collection.

We calibrated the maximum teleport and undo distance as \SI{10}{\meter}. For \ivMovementVisualizationLvlDisc{}, we chose the default values of SteamVR (\SI{0.2}{\second}, fade to black and back) for both teleport and undo. For \ivMovementVisualizationLvlCont{}, we chose a motion speed of \SI{10}{\meter/\second}. Further, we implemented a study client to control the study. Using an external monitor, we could further monitor the participants' actions. The study client logged the dependent variables to CSV files.

We deployed the application to a Gaming Laptop with Intel Core i7-9750H CPU @ 2.60GHz, 16GB RAM, and an NVIDIA GeForce RTX 2070. The participants wore a HTC Vive Pro and interacted with the default HTC Vive controller in their dominant hand. The size of the calibrated tracking area was \SI[parse-numbers=false]{2.7 x 2}{\meter}. While participants were free to move, both \ivPositionUndo{} and \ivOrientationUndo{} effectively overwrote intermediate user movements since the last teleport. To preserve the consistency of the virtual  and the physical world, we did not include the hand position in the undo. Accordingly, the relative position of the hand to the user's perspective remained the same after undo.

\subsection{Procedure}

After welcoming the participants, we introduced them to the concept. Then, we asked them to fill out a consent form together with a demographics questionnaire. We then described to the participants the exact procedure of the experiment and their task in the 8-arm radial maze, as well as the two experimental phases. After the participants could ask questions and we were confident that their task was clear to them, we started the first condition. 

We told the participants the combination of \ivOrientationUndo{}, \ivPositionUndo{}, and \ivMovementVisualization{} and started the system. In the following, participants had \SI{2}{\minute} to acclimatize with the locomotion method before we started the actual task. To start the first phase, the system placed the participants in the center of the central room with 4 doors closed. Once ready, the participants started a visual timer (\SI{3}{\second}) by pulling the trigger button. When the timer expired, the \experimentPhaseOne{} began. After participants collected the fourth coin, the remaining 4 doors opened without further cue. Since the participants were in a T-side arm at this point, this happened invisibly. Immediately after and without pause, the \experimentPhaseTwo{} started, in which the participants collected the remaining 4 coins. The system then enforced a \SI{1}{\minute} pause before the first repetition followed the described procedure.

After completing all two  repetitions of the condition, we asked participants to remove the VR goggles and complete the questionnaires on a tablet. We enforced a \SI{5}{min} break before starting the next condition. During this break, we asked the participants for further qualitative feedback in a semi-structured interview. Each experiment took about 100 minutes per participant. All participants and the investigator were vaccinated and anti-gen tested on the same day. Only the investigator and the participant were in the room at any given time. The investigator and participants wore medical face masks throughout the experiment. We disinfected all touched surfaces between the participants and ventilated the room for 30 minutes.  Our institutional ethics board reviewed and approved the study design.

\subsection{Participants}

We recruited 24 participants (5 identified as female, 19 as male) aged between 20 and 34
($\mu = 26.5$, $\sigma = 3.72$) from our university. 3 participants reported that they were first-time \ac{VR} users, 15 reported that they had used \ac{VR} before, and 6 reported that they were regular \ac{VR} users. Participants received compensation of around 15\$ in local currency.

\subsection{Analysis}
\label{sec:methodology:analysis}

We performed 3-way repeated measures (RM) ANOVAs with the \ivOrientationUndo{}, \ivPositionUndo{}, and \ivMovementVisualization{} as factors. For this, we first tested the data for violations of normality and sphericity assumptions using Shapiro-Wilk's and Mauchly's tests, respectively. If the assumption of normality was violated, we performed a non-parametric analysis. If the assumption of sphericity was violated, we corrected the tests using the Greenhouse-Geisser method and report the \gge{}. When the (RM) ANOVAs reported significant effects, we applied Bonferroni-corrected t-tests for post-hoc analysis. For the multi-factorial analysis of non-parametric data, such as the Likert questionnaires, we performed an \ac{ART} as proposed by \citet{Wobbrock2011} and applied the ART-C procedure as proposed by \citet{Elkin2021} for post-hoc analysis. Further, we report the generalized eta-squared \ges{} as an estimate of the effect size. As suggested by \citet{Bakeman2005}, we classify these effect sizes using Cohen's suggestions~\cite{Cohen1988} as small ($>.0099$), medium ($>.0588$), or large ($>.1379$). For count data, such as the number of teleports and errors, we fitted Poisson regression models and applied Type III Wald chi-square tests for significance testing.

\section{Results}
\label{sec:results}

In the following section, we report the results structured around the dependent variables described in \cref{sec:methodology}.

\subsection{Coin Collection Time}

\begin{figure*}[t!]
	
	\centering
	\includegraphics[width=\textwidth]{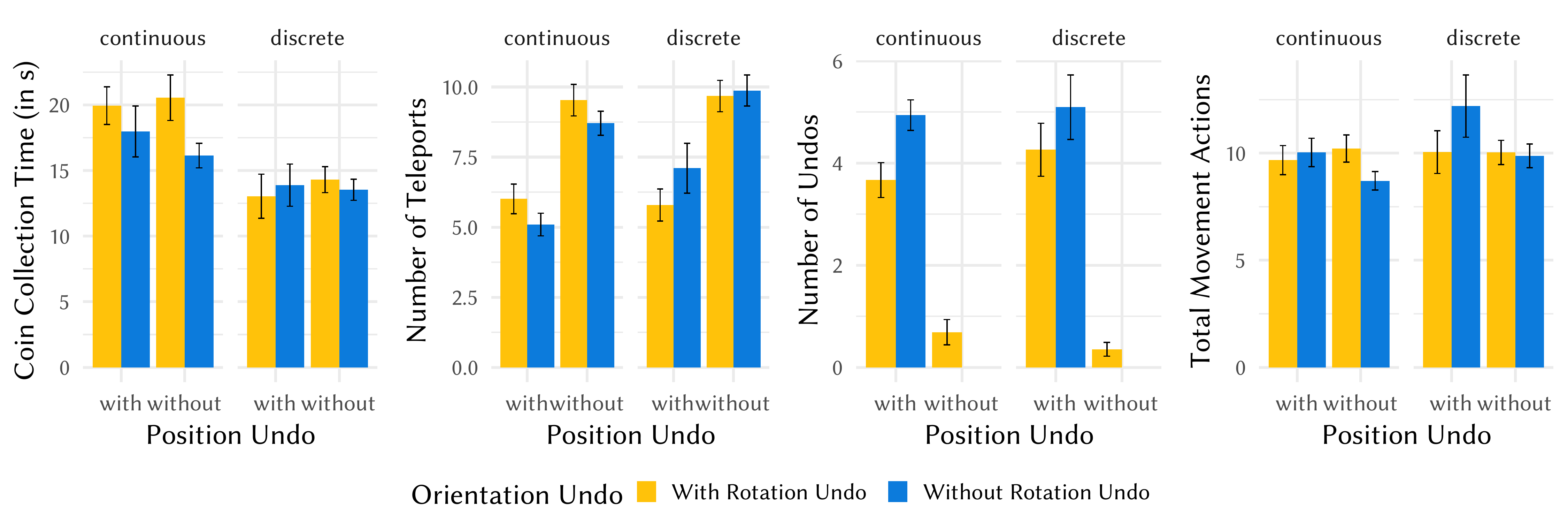}\hfill
	\vspace{-1em}
	\begin{minipage}[t]{.25\linewidth}
		\centering
		\subcaption{Coin Collection Time}\label{fig:undoport:results:coin_collection}
	\end{minipage}%
	\begin{minipage}[t]{.25\linewidth}
		\centering
		\subcaption{Number of Teleports}\label{fig:undoport:results:teleports}
	\end{minipage}%
	\begin{minipage}[t]{.25\linewidth}
	\centering
	\subcaption{Number of Undos}\label{fig:undoport:results:undos}
	\end{minipage}%
	\begin{minipage}[t]{.25\linewidth}
		\centering
		\subcaption{Total Movement Actions}\label{fig:undoport:results:movement_actions}
	\end{minipage}%
	
	\caption{\dvCollectionTime{} (a), \dvNrOfTeleports{} (b), \textsc{number of undos} (c) and \dvTotalMovementActions{} (d). Error bars depict the standard error.}
	\Description[Bar plots depicting the results of the experiment.]{Bar plots of the results of the experiment regarding the a) coin-collection time, b) number of teleports, c) number of undos and d) total movement actions.}
	
\end{figure*}

To assess the efficiency of participants, we measured the time needed to collect a coin. We found significantly shorter coin-collection times for \ivMovementVisualizationLvlDisc{} compared to \ivMovementVisualizationLvlCont{} with measured coin collection times ranging from \valSi{13.0}{9.41}{s} (both, discrete) to \valSi{20.5}{12.3}{s} (orientation only, continuous), see \cref{fig:undoport:results:coin_collection}.

We found a significant (\ano{1}{23}{25.42}{<.001}) influence of the \ivMovementVisualization{} with a \efETAsquared{0.12} effect size. Post-hoc tests confirmed significantly ($p<.001$) shorter coin-collection times for \ivMovementVisualizationLvlDisc{} (\valSi{13.7}{8.21}{s}) compared to \ivMovementVisualizationLvlCont{} (\valSi{18.7}{10.7}{s}). We could not find significant main effects of the \ivPositionUndo{} (\ano{1}{23}{.83}{>.05})  or the \ivOrientationUndo{} (\ano{1}{23}{2.35}{>.05}) nor interaction effects.

To exclude the influence of the different speeds in the two visualizations, we additionally analyzed the coin-collection time with the time for the actual teleports removed. We found coin-collection times ranging from \valSi{10.7}{6.8}{s} (no undo, continuous) to \valSi{14.4}{13.3}{s} (orientation only, continuous). We could not find significant main effects of \ivMovementVisualization{} (\ano{1}{23}{0.57}{>.05}), \ivPositionUndo{} (\ano{1}{23}{1.84}{>.05}) or \ivOrientationUndo{} (\ano{1}{23}{4.22}{>.05}) nor any interaction effects.

\subsection{Number of Teleports}

As another measurement of efficiency, we measured the number of teleports used to reach a coin. We found significantly higher numbers of teleports \ivPositionUndoLvlNo{} and \ivOrientationUndoLvlYes{} with mean numbers of teleports ranging from \val{5.09}{3.96} (positioin only, continuous) to \val{9.88}{5.42} (no undo, discrete), see \cref{fig:undoport:results:teleports}.

The analysis revealed a significant (\chisq{1}{75.37}{<.001}) main effect for the \ivPositionUndo{}. Post-hoc tests confirmed significantly ($p<.001$) higher numbers of teleports for \ivPositionUndoLvlNo{} (\val{9.45}{5.20}) compared to \ivPositionUndoLvlYes{} (\val{6.0}{6.14}).
Further, we found a significant (\chisq{1}{7.26}{<0.01}) main effect of the \ivOrientationUndo{} on the number of teleports. Post-hoc tests confirmed significantly ($p<.01$) higher numbers of teleports \ivOrientationUndoLvlYes{} (\val{7.75}{5.74}) compared to \ivOrientationUndoLvlNo{} (\val{7.70}{6.14}). We could not find a significant main effect for the \ivMovementVisualization{} (\chisq{1}{0.39}{>.05}).

Besides the main effects, we found a significant (\chisq{1}{19.44}{<.001}) interaction effect between \ivOrientationUndo{} and \ivMovementVisualization{}. While we could not find a difference in the number of teleports between \ivMovementVisualizationLvlDisc{} and \ivMovementVisualizationLvlCont{} for \ivOrientationUndoLvlYes{} (\val{7.73}{5.86} and \val{7.77 }{5.63}, $p>.05$), there was a significant ($p<.001$) difference for \ivOrientationUndoLvlNo{} (\ivMovementVisualizationLvlDisc{}: \val{8.49}{7.38} \ivMovementVisualizationLvlCont{}: \val{6.90}{4.46}).

\subsection{Number of Undo Actions}

To gain a deeper understanding of the usage of undo actions to reach a target, we analyzed the number of undo actions used to reach a target. We found that the type of undo support available had the strongest impact on the usage, with orientation-only support rarely used. We found a wide spread of usage numbers, ranging from \val{0.35}{1.33} (orientation only, discrete) to \val{5.09}{6.19} (position only, discrete), see \cref{fig:undoport:results:undos}.

For the analysis, we removed the data for the no undo conditions, considering the undo types (position-only, orientation-only, and both) as levels of a single factor. We found a significant (\chisq{2}{226.88}{<.001}) main effect on the number of undo actions. Post-hoc tests confirmed significant differences between all groups (orientation-only: \val{0.5}{1.97}, both: \val{3.96}{4.31} and position-only: \val{5.02}{4.83}, all $p<.001$). Further, the analysis revealed a significant (\chisq{1}{4.28}{<.05}) main effect of the \ivMovementVisualization{}. Post-hoc tests indicated significantly higher numbers of undo actions for discrete (\val{3.24}{5.11}) compared to continuous (\val{3.10}{3.42}).

Finally, we found a significant (\chisq{2}{13.45}{<.01}) interaction effect between undo type and \ivMovementVisualization{}. The analysis did not indicate significant differences between \ivMovementVisualizationLvlDisc{} and \ivMovementVisualizationLvlCont{} for the position-only and both conditions. For the orientation-only conditions, however, the analysis showed significantly higher usage for \ivMovementVisualizationLvlCont{} (\val{0.68}{2.44}) compared to the \ivMovementVisualizationLvlDisc{} (\val{0.35}{1.33}), $p<.05$. Nevertheless, the usage was still significantly lower compared to all other combinations (all $p<.001$).

\subsection{Streak Length}
 
We analyzed the number of successive actions of the same movement type (teleport and undo) as the streak length. We found comparable streaks between teleport and undo for the position-only and both conditions. For orientation-only, we found significantly shorter streak lengths for undo. Overall, we found streak lengths ranging from \val{3.77}{2.56} (positioin only, continuous) to \val{9.51}{2.79} (no undo, discrete) for teleport. For undo, we found streak lengths ranging from \val{2.88}{2.17} (orientation only, discrete) to \val{4.63}{2.45} (positioin only, continuous).

We excluded the conditions with no undo support, as the streak length would total the number of teleports. We found a significant (\chisq{3}{78.47}{<.001}) main effect of the available undo support. While we found longer mean streak lengths for orientation-only conditions (\val{7.90}{3.64}) compared to both other movement types (position only: \val{4.59}{4.34}, both: \val{4.21}{2.59}), post-hoc tests did not confirm significant differences.

Further, we found a significant (\chisq{2}{28.99}{<.001}) interaction effect between the available undo support and the movement type (i.e., teleport and undo). We could not find significant differences between the streak lengths between teleport and undo for the position-only (teleport: \val{4.64}{4.39}, undo: \val{4.54}{4.34}) and both (teleport: \val{4.67}{2.94}, undo: \val{3.71}{2.06}) conditions. For the orientation-only conditions, however, we found significantly longer teleport streaks (\val{9.13}{2.62}) compared to the undo streaks (\val{3.00}{3.01}).

\subsection{Total Movement Actions}

We further analyzed the number of total movement actions as the sum of teleport and undo movements used to reach a coin. The analysis indicated no main effects but showed interaction effects between \ivPositionUndo{} and \ivOrientationUndo{} and \ivMovementVisualization{} and \ivOrientationUndo{}, respectively, which we detail below. We found mean numbers ranging from \val{8.71}{4.22} (no undo, continuous) to \val{12.2}{14.2} (position only, discrete), see \cref{fig:undoport:results:movement_actions}.

While the analysis did not indicate any main effects (\ivMovementVisualization{}: \chisq{1}{.69}{>.05}, \ivPositionUndo{}: \chisq{1}{1.42}{>.05} and \ivOrientationUndo{}: \chisq{1}{.61}{>.05}), we found significant interaction effects between the independent variables. First, we found a significant (\chisq{1}{8.87}{<.01}) interaction effect between \ivPositionUndo{} and \ivOrientationUndo{}. While we could not find a difference in the number of total movement actions between both levels of \ivPositionUndo{} \ivOrientationUndoLvlYes{} (\ivPositionUndoLvlYes{}: \val{9.86}{8.33} \ivPositionUndoLvlNo{}: \val{10.1}{5.86}), we found significantly ($p<.001$) higher numbers of movement actions \ivPositionUndoLvlYes{} (\val{11.1}{11.1}) compared to \ivPositionUndoLvlNo{} (\val{9.29}{4.89}) \ivOrientationUndoLvlNo{}.

\begin{figure*}[t!]
	
	\centering
	\includegraphics[width=\textwidth]{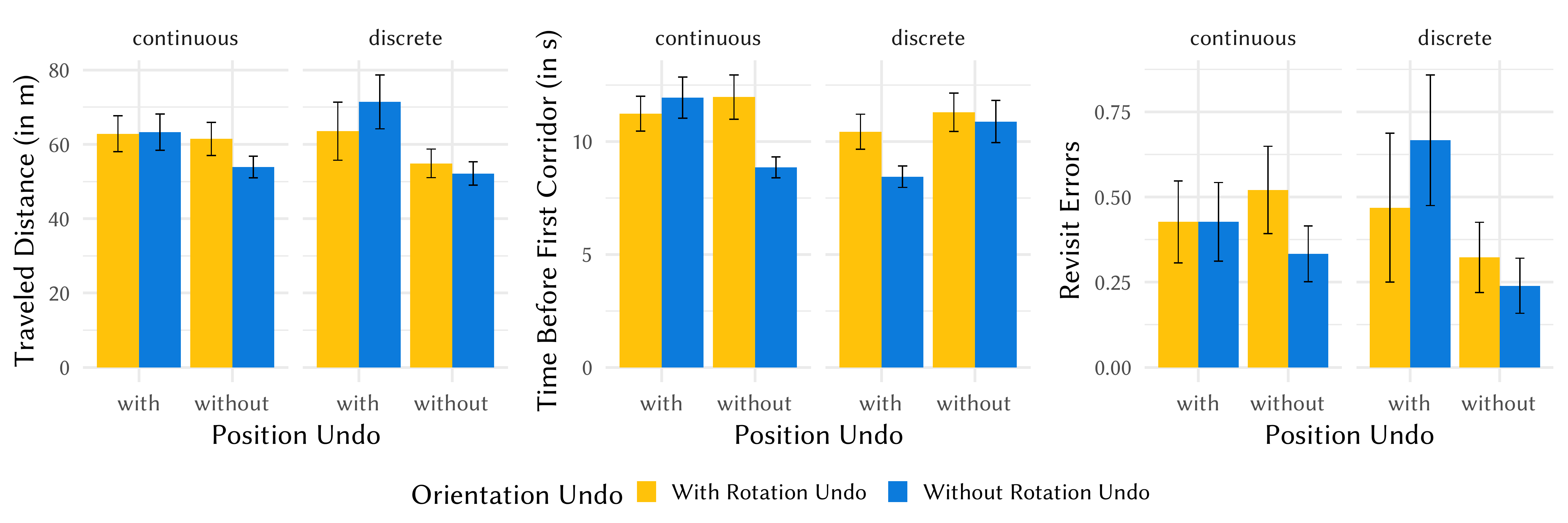}\hfill
	\vspace{-1em}
	\begin{minipage}[t]{.33\linewidth}
		\centering
		\subcaption{Traveled Distance}\label{fig:undoport:results:distance}
	\end{minipage}%
	\begin{minipage}[t]{.33\linewidth}
		\centering
		\subcaption{Time Before First Corridor}\label{fig:undoport:results:time_before_first_corridor}
	\end{minipage}%
	\begin{minipage}[t]{.33\linewidth}
		\centering
		\subcaption{Revisit Errors}\label{fig:undoport:results:revisit_errors}
	\end{minipage}%
\vspace*{-6pt}		
	\caption{\dvDistance{} (a), \dvTimeBeforeFirstCorridor{} (b) and \dvRevisitError{} (c). Error bars depict the standard error.}
	\Description[Bar plots depicting the results of the experiment.]{Bar plots of the results of the experiment regarding the a) traveled distance, b) time before entering the first corridor, and c) revisit errors.}
	
\end{figure*}

We found a significant (\chisq{1}{6.20}{<.05}) interaction effect between \ivMovementVisualization{} and \ivOrientationUndo{}. We we could not find differences for \ivMovementVisualizationLvlDisc{} (\val{10.0}{7.93}) and \ivMovementVisualizationLvlCont{} (\val{9.95}{6.40}) \ivOrientationUndoLvlYes{}, $p>.05$. \ivOrientationUndoLvlNo{}, however, we found significantly ($p<.001$) higher numbers for \ivMovementVisualizationLvlDisc{} (\val{11.0}{10.8}) compared to \ivMovementVisualizationLvlCont{} (\val{9.37}{5.51}).

\subsection{Traveled Distance}

We analyzed the traveled distance as another measure of efficiency.We found distances ranging from \valSi{52.2}{30.6}{m} (no undo, discrete) to \valSi{71.4}{71.0}{m} (position only, discrete), see \cref{fig:undoport:results:distance}. The analysis revealed a significant (\ano{1}{23}{7.69}{<.05}) main effect of the \ivPositionUndo{} on the traveled distance with a \efETAsquared{0.04} effect size. Post-hoc tests confirmed significantly ($p<.05$) higher traveled distances \ivPositionUndoLvlYes{} (\valSi{65.3}{62.0}{m}) compared to \ivPositionUndoLvlNo{} (\valSi{55.6}{35.6}{m}). We could not find other main (\ivMovementVisualization{}: \ano{1}{23}{0.24}{>.05}, \ivOrientationUndo{}: \ano{1}{23}{0.00}{>.05}) or interaction effects.

\subsection{Time Before First Corridor}

We measured the time before entering the first corridor to understand how closely they tried to memorize the surroundings. The analysis indicated an interaction effect between \ivMovementVisualization{} and \ivPositionUndo{} with lower times for the discrete visualization \ivPositionUndoLvlYes{}. Overall, we found times ranging from \valSi{8.44}{4.65}{s} (position only, discrete) to \valSi{12.0}{9.60}{s} (orientation only, continuous), see \cref{fig:undoport:results:time_before_first_corridor}.

While the analysis did not reveal any main effects (\ivMovementVisualization{}:  \ano{1}{23}{1.34}{>.05}, \ivPositionUndo{}: \ano{1}{23}{0.23}{>.05}, \ivOrientationUndo{}: \ano{1}{23}{1.40}{>.05}), we found an interaction effect.

The analysis revealed a significant (\ano{1}{23}{12.50}{<.01}) interaction effect between \ivMovementVisualization{} and \ivPositionUndo{} with a \efETAsquared{0.006} effect size. For \ivPositionUndoLvlNo{}, the analysis did not indicate a significant difference between both levels of \ivMovementVisualization{} (\ivMovementVisualizationLvlCont{}: \valSi{10.4}{7.65}{s}, \ivMovementVisualizationLvlDisc{}: \valSi{11.10}{8.71}{s}). For \ivPositionUndoLvlYes{}, however, we found a more pronounced, yet not significant, difference with lower times for discrete (\ivMovementVisualizationLvlCont{}: \valSi{11.6}{8.26}{s}, \ivMovementVisualizationLvlDisc{}: \valSi{9.44}{6.36}{s}).

\subsection{Revisit Error}

We logged the number of revisits to previously explored corridors as a measure of (dis-) orientation. We found significantly higher numbers of errors \ivPositionUndoLvlYes{}. Further, we found interaction effects between the independent variables, which we detail below. Overall, we found error rates per collected coin ranging from \val{0.24}{0.79} (no undo, discrete) to \val{0.67}{1.88} (position only, discrete), see \cref{fig:undoport:results:revisit_errors}.

The analysis indicated a significant (\chisq{1}{7.89}{<.01}) main effect of the \ivPositionUndo{} on the number of errors. Post-hoc tests confirmed significantly ($p<.01$) higher numbers of errors for \ivPositionUndoLvlYes{} (\val{0.50}{1.64}) compared to \ivPositionUndoLvlNo{} (\val{0.35}{0.99}). We could not find significant main effects for \ivMovementVisualization{} (\chisq{1}{0.01}{>.05}) or \ivOrientationUndo{} (\chisq{1}{0.14}{>.05}).

We found a significant (\chisq{1}{5.67}{<.05}) interaction effect between \ivPositionUndo{} and \ivOrientationUndo{}. For \ivOrientationUndoLvlYes{}, we found no significant difference \ivPositionUndoLvlYes{} (\val{0.45}{1.72}) and \ivPositionUndoLvlNo{} (\val{0.42}{1.14}), $p>.05$. For \ivOrientationUndoLvlNo{}, however, we found significantly ($p<.001$) higher numbers for \ivPositionUndoLvlYes{} (\val{0.55}{1.55}) compared to \ivPositionUndoLvlNo{} (\val{0.29}{0.80}).

Finally, we found a significant (\chisq{1}{8.56}{<.01}) interaction effect between \ivPositionUndo{} and \ivMovementVisualization{}. For the \ivMovementVisualizationLvlCont{} conditions, there was no significant difference between both levels of \ivPositionUndo{} (\ivPositionUndoLvlYes{}: \val{0.43}{1.15}, \ivPositionUndoLvlNo{}: \val{0.423}{1.06}), $p>.05$. For the \ivMovementVisualizationLvlDisc{} conditions, however, we found significantly higher numbers of errors ($p<.001$) for \ivPositionUndoLvlYes{} (\val{0.57}{2.01}) compared to \ivPositionUndoLvlNo{} (\val{0.28}{0.91}).

\subsection{\acl{TLX}}

\begin{figure*}[t!]
	
	\centering
	\includegraphics[width=\textwidth]{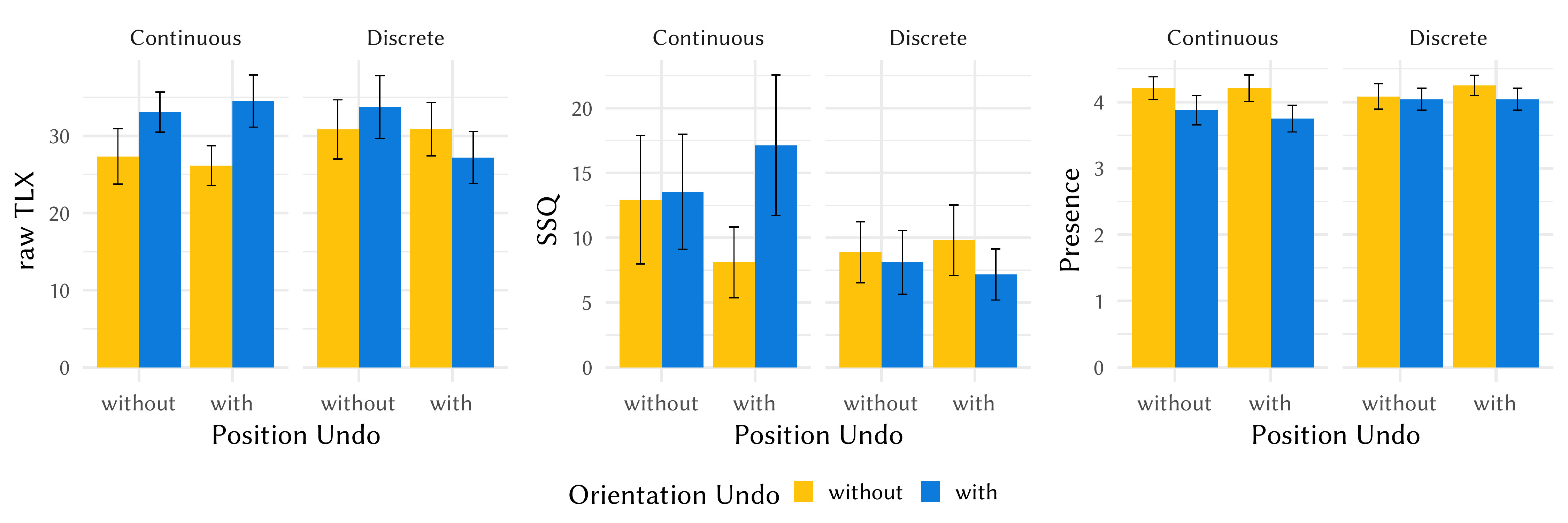}\hfill
	\vspace{-1em}
	\begin{minipage}[t]{.33\linewidth}
		\centering
		\subcaption{\acl{TLX}}\label{fig:undoport:results:tlx}
	\end{minipage}%
	\begin{minipage}[t]{.33\linewidth}
		\centering
		\subcaption{\acl{SSQ}}\label{fig:undoport:results:ssq}
	\end{minipage}%
	\begin{minipage}[t]{.33\linewidth}
		\centering
		\subcaption{Presence}\label{fig:undoport:results:presence}
	\end{minipage}%
\vspace*{-6pt}	
	\caption{The mean results of the (a) \dvTLX{}, (b) \dvSSQ{} and (c) \dvPresence{}. All error bars depict the standard error.}
	\Description[Bar plots depicting the results of the experiment.]{Bar plots of the results of the experiment regarding the a) NASA TLX, b) simulator sickness questionnaire, and c) presence.}
	\vspace*{-6pt}	
\end{figure*}

We assessed the \ac{TLX} as a measure of the perceived workload. We found significantly higher \ac{TLX} values for \ivOrientationUndoLvlYes{} with aggregated raw values ranging from \val{26.1}{12.6} (positioin only, continuous) to \val{34.5}{16.6} (both, continuous), see \cref{fig:undoport:results:tlx}.

The analysis indicated a significant (\ano{1}{23}{4.64}{<.05}) main effect of \ivOrientationUndo{} on the \ac{TLX}. Post-hoc tests confirmed significantly ($p<.05$) higher \ac{TLX} values for \ivOrientationUndoLvlYes{} (\val{32.1}{16.6}) compared to \ivOrientationUndoLvlNo{} (\val{28.8}{16.5}). We could not find further main (\ivMovementVisualization{}: \ano{1}{23}{0.03}{>.05}, \ivPositionUndo{}: \ano{1}{23}{0.56}{>.05}) or interaction effects.

\subsection{\acl{SSQ}}

We assessed the \ac{SSQ} to assess the influences of our proposed interaction techniques on the\break experienced simulator sickness. We found significantly higher \ac{SSQ} values for \ivOrientationUndoLvlYes{} as well as an interaction effect between \ivOrientationUndo{} and \ivMovementVisualization{}, which we detail below. Overall, we found mean values ranging from \val{7.17}{9.67} (both, discrete) to \val{17.1}{26.6} (both, continuous), see \cref{fig:undoport:results:ssq}.

Shapiro-Wilk's test indicated a violation of the assumption of normality of the residuals. Therefore, we analyzed the data using the \acl{ART} approach as outlined in \cref{sec:methodology:analysis}. The \ac{ART} ANOVA indicated a significant (\ano{1}{23}{5.95}{<.05}) main effect for \ivOrientationUndo{} with a \efETAsquared{0.21} effect size. Post-hoc tests confirmed significantly ($p<.05$) higher \ac{SSQ} scores for \ivOrientationUndoLvlYes{} (\val{11.5}{19.0})  compared to \ivOrientationUndoLvlNo{} (\val{9.93}{16.3}). We could not find further main effects (\ivMovementVisualization{}: \ano{1}{23}{0.24}{>.05}, \ivPositionUndo{}: \ano{1}{23}{2.18}{>.05}).

Further, we found a significant (\ano{1}{23}{11.62}{<.01}) interaction effect between \ivOrientationUndo{} and \ivMovementVisualization{}. We did not find significant differences in the \ac{SSQ} scores for both levels of \ivOrientationUndo{} for the \ivMovementVisualizationLvlDisc{} visualization (\ivOrientationUndoLvlNo{}: \val{9.35}{12.3}, \ivOrientationUndoLvlYes{}: \val{7.64}{10.8}), $p>.05$. For \ivMovementVisualizationLvlCont{}, however, we found significantly ($<.05$) higher \ac{SSQ} scores for \ivOrientationUndoLvlYes{} (\val{15.3}{24.1}) compared to \ivOrientationUndoLvlNo{} (\val{10.5}{19.5}).

\subsection{Presence}

We assessed the participants' feeling of presence through the answer to the question \enquote{In the computer-generated world I had a sense of \enquote{being there}} on a 7-point Likert scale (\enquote{1: not at all} $\ldots$ \enquote{7: very much}). We found significantly higher presence ratings for \ivOrientationUndoLvlNo{} with answers ranging from \val{3.75}{0.99} (both, continuous) to \val{4.25}{0.74} (position only, discrete), see \cref{fig:undoport:results:presence}.

The \ac{ART} ANOVA revealed a significant (\ano{1}{23}{20.88}{<.001}) main effect for the \ivOrientationUndo{} on the participants'\break ratings of the presence statement with a \efETAsquared{0.48} effect size. Post-hoc tests confirmed significantly higher presence ratings for \ivOrientationUndoLvlNo{} (\val{4.19}{0.86}) compared to \ivOrientationUndoLvlYes{} (\val{3.93}{0.92}). We could not find any other significant main (\ivMovementVisualization{}: \ano{1}{23}{0.74}{>0.5}, \ivPositionUndo{}: \ano{1}{23}{0.02}{>.05}) or interaction effects.

\subsection{Custom Questionnaire}

\begin{figure*}[ht!]
	
	\centering
	\includegraphics[width=\textwidth]{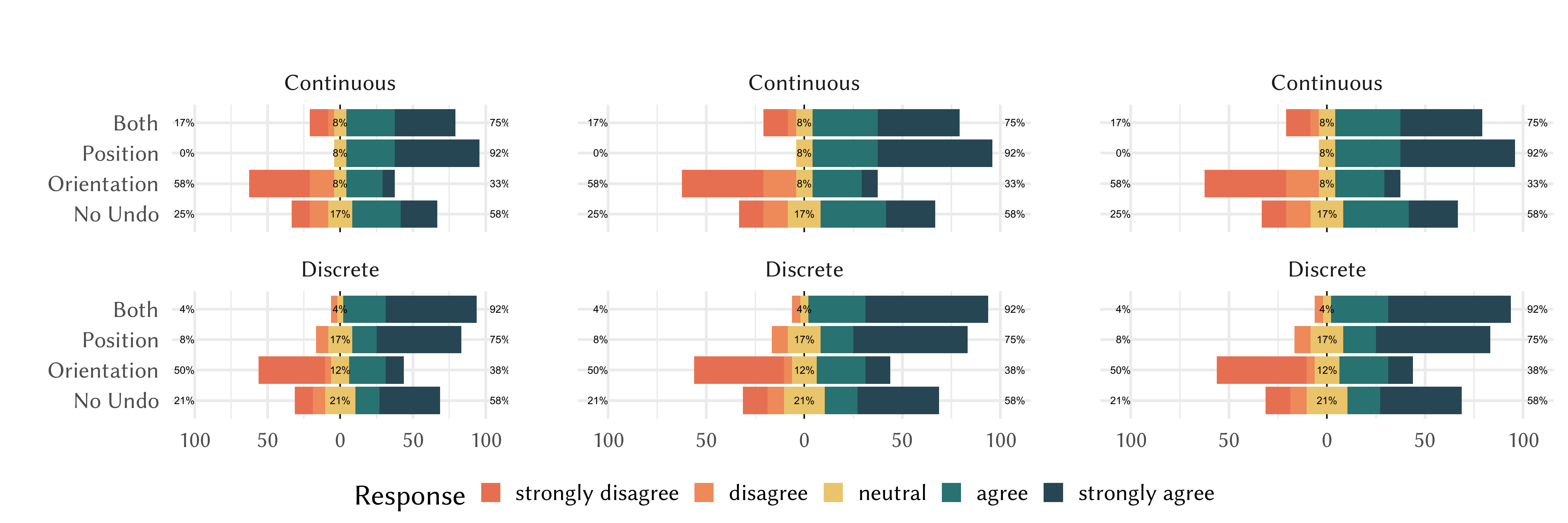}\hfill
	\vspace{-1em}
	\begin{minipage}[t]{.33\linewidth}
		\centering
		\subcaption{Helpfulness}\label{fig:undoport:results:likert_q1}
	\end{minipage}%
	\begin{minipage}[t]{.33\linewidth}
		\centering
		\subcaption{Convenience}\label{fig:undoport:results:likert_q2}
	\end{minipage}%
	\begin{minipage}[t]{.33\linewidth}
		\centering
		\subcaption{Orientation Problems}\label{fig:undoport:results:likert_q3}
	\end{minipage}%
	
	\caption{The participants' answers to our questions on a 5-point Likert scale regarding the perceived (a) helpfulness, (b) convenience and (c) orientation problems. For full questions, please refer to the text.}
	\Description[Plots depicting the participants' answers on a 5-point Likert scale.]{Plots depicting the participants' answers to three questions on a 5-point Likert scale regarding a) helpfulness, b) convenience, and c) perceived orientation problems.}
	
\end{figure*}

As a last measure, participants answered three custom questions on a 5-point Likert scale. In the following section, we analyze their answers.

\paragraph{\enquote{The locomotion technique helped me complete my task.}}

We found a significant (\ano{1}{23}{73.80}{<.001}) influence of the \ivPositionUndo{} on the participants' answers with a \efETAsquared{0.09} effect size. Post-hoc tests confirmed significantly ($p<.001$) higher ratings for \ivPositionUndoLvlYes{} compared to \ivPositionUndoLvlNo{}. Further, we found a significant (\ano{1}{23}{12.33}{<.01}) influence of the \ivOrientationUndo{} with a \efETAsquared{0.21} effect size. Post-hoc tests confirmed significantly ($p<.01$) higher ratings for \ivOrientationUndoLvlNo{} compared to \ivOrientationUndoLvlYes{}. Finally, we found a significant (\ano{1}{23}{17.30}{<.001}) interaction effect between \ivPositionUndo{} and \ivOrientationUndo{}. The combination with \ivOrientationUndoLvlYes{} was rated significantly ($p<.001$) less helpful compared to \ivOrientationUndoLvlNo{} for \ivPositionUndoLvlNo{}. For \ivPositionUndoLvlYes{}, however, the effect was turned upside down, and participants rated the combination with \ivOrientationUndoLvlYes{} significantly ($p<.001$) more helpful compared to \ivOrientationUndoLvlNo{}. Figure \ref{fig:undoport:results:likert_q1} shows all answers from our participants.

\paragraph{\enquote{The locomotion technique was convenient to use.}}

\looseness-1We found significant main effects for all three independent variables (\ivPositionUndo{}: \ano{1}{23}{35.43}{<.001} with a \efETAsquared{0.08} effect size, \ivOrientationUndo{}: \ano{1}{23}{25.28}{<.001} with a \efETAsquared{0.21} effect size and \ivMovementVisualization{}: \ano{1}{23}{14.55}{<.001} with a \efETAsquared{0.01}). Post-hoc tests confirmed significantly (all $p<.001$) higher ratings for \ivPositionUndoLvlYes{}, \ivOrientationUndoLvlNo{} and \ivMovementVisualizationLvlDisc{} compared to their respective counterparts. Further, we found interaction effects. First, we found a significant (\ano{1}{23}{8.10}{<.01} interaction effect between \ivPositionUndo{} and \ivOrientationUndo{} with a \efETAsquared{0.13} effect size. For \ivPositionUndoLvlNo{}, we found that participants rated the convenience significantly lower for \ivOrientationUndoLvlYes{} compared to \ivOrientationUndoLvlNo{} ($p<.01$). For \ivPositionUndoLvlYes{}, however, participants found \ivOrientationUndoLvlYes{} significantly more convenient compared to \ivOrientationUndoLvlNo{} ($p<.001$). Finally, we found a significant (\ano{1}{23}{22.44}{<.001}) interaction effect between \ivOrientationUndo{} and \ivMovementVisualization{} with a \efETAsquared{0.34} effect size. For \ivMovementVisualizationLvlCont{}, \ivOrientationUndoLvlYes{} was rated significantly ($p<.001$) lower compared to \ivOrientationUndoLvlNo{}. For \ivMovementVisualizationLvlDisc{}, however, we found the opposite effect and better ratings for \ivOrientationUndoLvlYes{}. Yet, the difference was not significant ($p>.05$). Figure \ref{fig:undoport:results:likert_q2} shows all answers from our participants.

\paragraph{\enquote{I had problems orienting myself.}}

For the last question, the analysis did not indicate any main effects in the data (\ivPositionUndo{}: \ano{1}{23}{2.36}{>.05}, \ivOrientationUndo{}: \ano{1}{23}{2.62}{>.05} and \ivMovementVisualization{}: \ano{1}{23}{0.28}{>.05}). However, we found a significant (\ano{1}{23}{5.37}{<.05}) interaction effect between \ivOrientationUndo{} and \ivMovementVisualization{} with a \efETAsquared{0.34} effect size. For \ivMovementVisualizationLvlDisc{}, we participants reported significantly higher levels of orientation-loss for \ivOrientationUndoLvlNo{} compared to \ivOrientationUndoLvlYes{} ($p<.05$). For \ivMovementVisualizationLvlCont{}, however, we found that participants reported higher levels of orientation-loss for \ivOrientationUndoLvlYes{} compared to \ivOrientationUndoLvlNo{}. Yet, the difference was not significant ($p>.05$). \Cref{fig:undoport:results:likert_q3} shows all answers from our participants.

\subsection{Qualitative Feedback}

In general, our participants showed strong approval for the idea of reverting locomotion steps through undo actions. 

In particular, the \ivPositionUndo{} was positively received by 22 of the 24 participants. Asked for the reasons, participants reported that position undo was \pquote{convenient}{P5, P16} and \pquote{helpful}{P6} as there was \pquote{no need to turn]}{P4} which made it \pquote{quicker to navigate back [...] without having to physically turn around}{P3}. This helped to \pquote{[not] lose your orientation so easily.}{P21}. Further, participants commented that it is \pquote{nice to have when doing errors}{P11} 

Regarding the \ivOrientationUndo{}, participants' opinions were split, with 13 out of the 24 participants preferring to have orientation support. Participants described their experiences with orientation undo as \pquote{faster [than physically turning back]}{P6} and \pquote{helpful}{P5, P8} as \pquote{it helps to establish a familiar starting position}{P7}. In contrast, other participants reported that it \pquote{made me lose orientation}{P17} by causing \pquote{irritation in my sense of space}{P2}. Further, participants reported increased cybersickness as it \pquote{made motion sickness worse}{P11} and \pquote{just made me dizzy and feel disconnected}{P12}. To explain this mismatch between the positive and negative aspects, P18 explained that it depends on what level of \ivPositionUndo{} it was paired with: \pquote{[orientation support] messed up my orientation. Except when combined with position undo}{P18}. Other participants agreed as \pquote{orientation reset [...] without position reset felt [...] useless.}{P10} while it was considered \pquote{helpful}{P9, P10, P16} when used \pquote{together}{9} and in \pquote{combination}{P1, P10, P16, P18} with position undo.

The question about the preferred \ivMovementVisualization{} again showed a mixed picture, with a clear tendency towards \ivMovementVisualizationLvlDisc{} visualization. While 6 participants preferred \ivMovementVisualizationLvlCont{}, the other 18 saw advantages in the \ivMovementVisualizationLvlDisc{} visualization. Asked for the reasons for preferring the \ivMovementVisualizationLvlDisc{} visualization, participants explained that it felt \pquote{faster}{P1, P3, P5, P6, P8, P9, P10, P21, P22} and caused \pquote{less vertigo}{P3} and \pquote{less nausea}{P8, P22}. The \ivMovementVisualizationLvlCont{} visualization was found to provide \pquote{better orientation}{P2, P4} and a \pquote{greater immersion in the virtual world}{P7} which \pquote{helped [..] with orientation}{P15}. As a possible reason, P20 explained that \enquote{you can see the route you are traveling}.

\section{Discussion}
\label{sec:discussion}

The results of our controlled experiment suggest that undo actions provide a viable addition to \teleport{}. We found that undo actions can increase the efficiency of participants when locomoting in virtual environments and received very favorable feedback from our participants. However, we also found a negative impact on the participants' ability to orientate themselves. In the following section, we discuss the results in relation to our research questions.

\subsection{Position Undo Allows for Faster Travel but Increases Errors}

In our analysis, we found significantly higher numbers of errors, which required participants to perform a higher number of teleports and, subsequently, higher traveled distances to collect a coin \ivPositionUndoLvlYes{}. Surprisingly, this increased movement of the participants was not reflected in the time used to collect a coin. We found no significant effect of position undo on time required and even the lowest average time measured for a condition with position undo (both, discrete).

We attribute this finding to a combination of several effects. First, the cost (in terms of distance covered per time) for undo actions is lower than for standard locomotion actions. This is explained by the ability to travel a distance comparable to a normal teleport with a simple button press, requiring no prior physical body rotation and no targeting. Due to the radial maze design of the task, about half of all distances in the \ivPositionUndoLvlYes{} conditions could be covered by jumping back as participants had to return to the central room on their way to the next coin. Second, while we found no negative influence on participants' orientation in their subjective self-assessment that could explain the increased error rates, we found evidence that the lower cost for undo actions contributed to this. The participant's task was to collect the coins as quickly as possible. Using the radial arm maze task, we sought to increase the time cost of a circular search to encourage participants to rely on their orientation in the search rather than visiting all paths sequentially. Looking at our results, however, we hypothesize that by reducing the effort to travel back through undo actions, participants paid less attention to preventing errors and perceived it as more efficient to check the corridors one after the other. This increased the error rate but resulted in comparable times due to the inherently faster movement. The data from our participants supports this interpretation, as they rated \ivPositionUndoLvlYes{} as significantly more convenient and helpful in completing the task. In addition to the quantitative results, the qualitative feedback supports this interpretation as participants reported that undo support helped resolve errors.

\subsection{Orientation Undo Alone Has a Negative Effect but Can Enhance the Positive Characteristics of Position Undo.}

As position undo, \ivOrientationUndo{} significantly increased the number of errors and, consequently, the number of teleports and the distance traveled per coin. But, again, these increased travel distances did not result in increased coin-collection times. Further, we found a negative effect on the \ac{TLX}, the \ac{SSQ}, and the perceived presence. However, these individual results do not provide the complete picture. In combination with \ivPositionUndoLvlYes{}, orientation undo reduced the number of errors and was rated significantly more convenient and helpful than without orientation undo. This finding is reinforced by the qualitative feedback, where most participants preferred orientation undo to no orientation undo, but only in combination with position undo. We, therefore, attribute many of the individual negative results of orientation undo to the poor performance of the technique without position undo.

The good performance of the combination of position and orientation undo appears intuitively understandable, given that both dimensions are reset in one step, resulting in the lowest cognitive load of the discrete techniques. Surprisingly, however, position-only undo without orientation worked comparably well across all measures, whereas orientation-only undo received poor ratings. Further, it was used much less: While participants used a similar number of teleports and undos in position-only and both conditions and the streak length also showed no differences, undo actions were used only very rarely in orientation-only conditions. While we do not have a conclusive explanation for this, we attribute this effect to the unique properties of motion in \ac{VR}. Changing position in VR involves a relatively large amount of effort aiming with the controller. This intermediate step of aiming is omitted with positions undo. For orientation changes, however, users only need to turn their heads, which implies a lower effort. Accordingly, we hypothesize that our participants were more likely to accept the potential drawbacks of increased cognitive load from undo actions with position undo, while the benefits for orientation resets alone were too small to offset the drawbacks. Further work is needed in this area to conclude on these questions.

\subsection{Prefer Discrete over Continuous Visualization}

For the \ivMovementVisualization{}, the discrete visualization in our experiment showed clear advantages over the continuous visualization in many ways. For example, the discrete visualization led to lower coin-collection times, reduced cybersickness, and was clearly preferred by the participants in both quantitative and qualitative feedback.

In particular, the continuous movement visualization was negatively evaluated with orientation undo. We attribute this effect to the increased mismatch between the virtual camera rotation and the lack of physical head rotation, increasing cybersickness and potentially affecting the other measures. This is supported by significantly higher levels of reported orientation loss with orientation undo in the continuous visualization conditions and predominantly negative feedback in the quantitative and qualitative feedback by the participants.

We acknowledge that some of the limitations found for the continuous visualization might be based on the implementation details, such as the duration, and other implementations might yield other results. However, we are confident that our results provide valuable insights into the design space of both discrete and continuous movement visualizations for undo actions.

\subsection{How to Undo?}

Taken together, our results support the use of undo actions compared to state-of-the-art, which we explored as a baseline (no undo, discrete) in the study. Further, while continuous movement and the single use of orientation undo did not translate into improvements in quantitative and qualitative data, we found strong support for position undo.

Our participants perceived position undo positively, and the quantitative data confirmed fast travel speeds (as distance traveled per time) with and without orientation support. Further, our data show that the undo options were frequently used in the position-only condition and in both conditions, although we left it up to participants to decide how they wanted to move. We found that the different movement options were often used in a series of 3-5 actions. This finding is consistent with our observations from the experiment: Participants teleported to a target required, on average, 3-5 teleports as they did not use the maximum teleport distance. After collecting the coin (or discovering a mistake), participants used a quick succession of position undo actions to return to the starting point, if available in the condition. The question of which of the options was performing and perceived better seemed to be largely based on user preferences in our controlled experiment. 

Therefore, we propose to provide users with the option to revert their movements with position and (optionally) orientation in future \ac{VR} experiences. However, a deeper investigation of the higher number of errors is needed in the future to explore whether the found rising number is an artifact introduced by the study task design or a consequence of the interaction technique.

\section{Limitations and Future Work}
\label{sec:limitions}

We are convinced that the presented concepts and results of our evaluation provide valuable insights and guidelines for the future use of undo actions for locomotion in \ac{VR} environments. However, our experiment's design and results impose some limitations and directions for future work, which we discuss in the following.

\subsection{Ecological Validity and Real-World Applicability}

In this paper, we contributed an experiment that deliberately adopted a highly artificial and reduced virtual environment and task. We chose this approach to exclude external influencing factors (for example, landmarks in the world) and to assess the pure effect of the presented locomotion techniques on efficiency, orientation, and user experience. In particular, we wanted to measure the effects on users' orientation abilities without landmarks in the \ac{VR} scene. These would have turned the task from a pure orientation task to a memory task.

In realistic \ac{VR} scenarios, however, users will find spatial cues as landmarks in the virtual world. Previous work has shown that such spatial cues greatly impact users' orientation ability~\cite{Harris2011}. Such cues can, thus, help to mitigate the negative effects of some of the interaction techniques presented here on orientation ability while keeping the positive effect on efficiency. Future work in this area is needed to assess the influence of spatial cues and their interaction with the techniques presented. However, we are confident that our work can serve as a baseline for this.

\subsection{Generalizability to Other Locomotion Techniques}

In this work, we explored undo actions for virtual locomotion using various extensions to \teleport{}. We chose this approach to study the impact of our extensions on the most commonly used interaction technique.

However, in recent years, research has brought forth a wide variety of other artificial locomotion techniques (see \cref{sec:relatedwork:locomotion}), yielding different requirements and implications for the inclusion and design of undo actions. Although we are confident that the benefits of undo actions demonstrated in this work can also be applied to these techniques, further work is needed to investigate the impact of undo actions on efficiency, orientation, and user experience in these scenarios.

\subsection{Undoing Time}

This work explored the use of locomotion undo actions, undoing spatial changes while moving through the virtual environment. However, the known mental model for undo actions, as users know it from interaction with computer systems, describes something different: Here, an undo reverses an action as if it had never happened~\cite{Yang1988, Abowd1992}. In the picture of interaction in a \ac{VR} world, not only the movement would be undone, but also the further actions in the world; in a way, it would be a rewinding of time. This rewinding of time has already been investigated for desktops~\cite{Shi2021a, Kleinman2020} and, recently, for \ac{VR} scenarios~\cite{Doma2022}.

While we find this approach highly intriguing and promising, we deliberately opted not to include a temporal undo in our design. This design decision is rooted in our work's specific intention to target the process of locomotion in \ac{VR}. Therefore, the simultaneous undoing of time and, thus, the users' actions in the scene would effectively prevent this from being used as a locomotion technique, as any action (e.g., collecting a coin) would be undone at the same time.

\subsection{Refinement of Undo Actions}

In our work, we found a negative influence of undo actions on the number of errors, which we assessed as a measure of the users' spatial orientation. For our experiment, we investigated continuous visualization of the locomotion process as a possible way to strengthen spatial understanding. However, we found negative influences of this visualization on participants' cybersickness and liking, yielding a clear advantage of the discrete visualization in our results.

We suggest investigating additional visualization techniques as possible alternatives to strengthen participants' sense of orientation while maintaining the benefits of undo actions. For example, a motion blur in the style of superimposed teleport images could provide an intermediate between discrete and continuous visualization. Also, adding a visual indication of where one lands when undoing - similar to the position indication provided by the teleport beam - could help users orient themselves in the scene. As another way to increase users' orientation, we suggest investigating an alternative implementation of orientation undo: In our experiment, we reset the orientation based on the gaze direction (i.e., the orientation of the \acp{HMD}). This could adversely affect the user's orientation ability since only the perspective is restored, not the body pose. An alternative implementation based on the forward vector of the body could alleviate this concern. However, it would require additional tracking hardware or rely on heuristics based on the position of the controller and head position, which would inherently introduce some uncertainty. Further work is needed to conclude on the best design for undo actions.

\section{Conclusion}
\label{sec:conclusion}

In this paper, we explored the effect of undo actions as an extension to \teleport{} on the participants' efficiency, orientation, and user experience. For this, we proposed eight variations of undo actions based on the availability of position and orientation undo and different movement visualizations. We compared the variations in a controlled experiment with 24 participants. We found promising results, indicating that undo action can provide users with an easy and fast option to skip travel times for returning to previously visited locations in \ac{VR}. However, our results indicate that undo actions can negatively influence the participant's spatial orientation in the virtual scene, calling for further research in this direction.

\begin{acks}
	Figure \ref{fig:undoport:ivs} uses material designed by \href{https://freepik.com/}{freepik}. We thank our anonymous reviewers for their valuable comments and suggestions and all participants who took part in our experiment. This work has been funded by the \grantsponsor{h2020}{European Union's Horizon 2020 research and innovation program}{http://ec.europa.eu/programmes/horizon2020/en} under grant agreement No.	\grantnum[https://www.humane-ai.eu/]{h2020}{952026}.
\end{acks}

\balance

\bibliographystyle{ACM-Reference-Format}

\end{document}